\documentclass[12pt]{article}
\usepackage{epsfig,amssymb,amsmath}
\newcommand{\Dlr}{\stackrel{\leftrightarrow}{D}}
\newcommand{\Dl}{\stackrel{\leftarrow}{D}}
\newcommand{\Dr}{\stackrel{\rightarrow}{D}}

\newcommand{\pslash}{\not{\hspace{-0.08cm}p}}

\newcommand{\be}{\begin{equation}}
\newcommand{\ee}{\end{equation}}

\newcommand{\bea}{\begin{eqnarray}}
\newcommand{\eea}{\end{eqnarray}}
\newcommand{\MS}{{\overline{MS}}}

\title{
\vspace{-2.5cm}
\flushleft{\normalsize DESY 05-029} \\
\vspace{-0.35cm}
{\normalsize Edinburgh 2005/03} \\
\vspace{-0.35cm}
{\normalsize Leipzig LU-ITP 2005/014} \\
\vspace{-0.35cm}
{\normalsize Liverpool LTH 648} \\
\vspace{0.35cm}
\centering{\Large \bf Renormalisation of one-link quark operators for
overlap fermions with L\"uscher-Weisz gauge action}\vspace*{-0.35cm}}
\author{\large R.~Horsley$^1$, H.~Perlt$^{2,3}$, P.E.L.~Rakow$^4$,
G.~Schierholz$^{5,6}$ and A.~Schiller$^3$ \\[0.75em]
-- QCDSF Collaboration -- \\[0.75em]
\normalsize
$^1$ School of Physics, University of Edinburgh, \\
\normalsize
Edinburgh EH9 3JZ, UK \\
\normalsize
$^2$ Institut f\"ur Theoretische Physik, Universit\"at Regensburg,\\
\normalsize
D-93040 Regensburg, Germany \\
\normalsize
$^3$ Institut f\"ur Theoretische Physik, Universit\"at Leipzig, \\
\normalsize
D-04109 Leipzig, Germany \\
\normalsize
$^4$ Theoretical Physics Division, Department of Mathematical Sciences,\\
\normalsize
University of Liverpool,\\
\normalsize
Liverpool L69 3BX, UK \\
\normalsize
$^5$ John von Neumann-Institut f\"ur Computing NIC,\\
\normalsize
Deutsches Elektronen-Synchrotron DESY,\\
\normalsize
D-15738 Zeuthen, Germany\\
\normalsize
$^6$ Deutsches Elektronen-Synchrotron DESY, \\
\normalsize
D-22603 Hamburg, Germany
}

\date{ }

\begin{document}

\maketitle

\begin{abstract}
We compute lattice renormalisation constants of one-link quark operators
({\it i.e.} operators with one covariant derivative)
for overlap fermions and L\"uscher-Weisz gauge action in one-loop perturbation
theory. Among others, such operators enter the calculation of moments
of polarised and unpolarised hadron structure functions.
Results are given for $\beta=8.45$, $\beta=8.0$ and mass parameter
$\rho=1.4$, which are commonly used in numerical simulations. We apply mean
field (tadpole) improvement to our results.
\end{abstract}

\section{Introduction}

In a recent publication~\cite{Horsley:2004mx} we have computed lattice
renormalisation constants of local bilinear quark operators for overlap
fermions and improved gauge actions in one-loop perturbation theory. Among the
actions we considered were the
Symanzik, L\"uscher-Weisz, Iwasaki and DBW2 gauge actions. The results were
given for a variety of $\rho$ parameters. Furthermore, we showed how to apply
mean field (tadpole) improvement to overlap fermions. In this letter we shall
extend our work to one-link bilinear quark operators. Operators of this kind
enter, for example, the calculation of moments of polarised and unpolarised
hadron structure functions. The present calculations are much more involved
than the previous ones, so that we shall restrict ourselves to the
L\"uscher-Weisz action, and to parameters actually being used in numerical
calculations.

The integral part of the overlap fermion action~\cite{overlap,n2,n}
\be
S_F=\bar{\psi}\left[\left(1-\frac{am}{2}\right) D_N + m\right]\psi \,,
\ee
$m$ being the mass of the quark, is the Neuberger-Dirac operator
\be
D_N = \frac{\rho}{a} \left( 1+ \frac{X}{\sqrt{X^\dagger X}}\right)\,,\;
X = D_W - \frac{\rho}{a}\,,
\label{over}
\ee
where $D_W$ is the Wilson-Dirac operator, and $\rho$ is a real parameter
corresponding to a negative mass term. At tree level  $0 < \rho < 2r$, where
$r$ is the Wilson parameter. We take $r=1$ and consider massless quarks.

Numerical simulations of overlap fermions are significantly more costly
than simulations of Wilson fermions. The cost of overlap fermions is
largely determined by the condition number of $X^\dagger X$. This number is
greatly reduced for improved gauge field actions~\cite{DeGrand:2002vu}. For
example, for the tadpole improved L\"uscher-Weisz action we found a reduction
factor of $\gtrsim 3$ compared to the Wilson gauge field
action~\cite{Galletly:2003vf}. The reason is that the L\"uscher-Weisz action
suppresses unphysical zero modes, sometimes called
dislocations~\cite{Gockeler:1989qg}. A reduction of the number of small modes
of $X^\dagger X$ appears also to result in an
improvement of the locality of the overlap operator~\cite{DeGrand:2002vu}.

We consider the
tadpole improved L\"uscher-Weisz
action~\cite{Luscher:1984xn,Luscher:1985zq,Alford:1995hw}
\be
\begin{split}
S_G & = \frac{6}{g^2} \,\,\Bigg[c_0
\sum_{\rm plaquette} \frac{1}{3}\, {\rm Re\, Tr\,}(1-U_{\rm plaquette})
\, + \, c_1 \sum_{\rm rectangle} \frac{1}{3}\,{\rm Re \, Tr\,}(1- U_{\rm
rectangle}) \\
&\phantom{=}  + \, c_3 \sum_{\rm parallelogram} \frac{1}{3}\,{\rm Re \,Tr\,}(1-
U_{\rm parallelogram})\Bigg]\,,
\end{split}
\label{gluonaction}
\ee
where $U_{\rm plaquette}$ is the standard plaquette, $U_{\rm rectangle}$
denotes the loop of link matrices around the $1\times 2$ rectangle,
and $U_{\rm paralellogram}$ denotes the loop along the edges of the
three-dimensional cube. It is required that $c_0+ 8 c_1 + 8 c_3=1$ in the
limit $g\rightarrow 0$, in order to ensure the correct continuum limit.
We define
\be
\beta = \frac{6}{g^2}\, c_0\,.
\label{defbeta}
\ee
The remaining parameters are~\cite{Alford:1995hw}:
\be
\frac{c_1}{c_0}=-\frac{(1+0.4805\,\alpha)}{20\,u_0^2}\,,\quad
\frac{c_3}{c_0}=-\frac{0.03325\,\alpha}{u_0^2}\,, \quad
\frac{1}{c_0}=1+8\left(\frac{c_1}{c_0}+\frac{c_3}{c_0}\right)\,,
\ee
where
\be
u_0=\left(\frac{1}{3}\, {\rm Tr}\,\langle U_{\rm
    plaquette}\rangle\right)^{\frac{1}{4}}\, , \quad
\alpha = -\frac{\log(u_0^4)}{3.06839}\,.
\label{u0}
\ee

The final results cannot be expressed in analytic form (as a function of $\beta$ and
$\rho$) anymore. We therefore have to make a choice. Here we consider two couplings,
 $\beta=8.45$ and $8.0$, at which we run Monte Carlo simulations at present~\cite{Galletly:2003vf,Gurtler:2004ac}. The corresponding values of
$c_1$ and $c_3$ are~\cite{Gattringer:2001jf}:
\be
\begin{tabular}{c|c|c|c}
  $\beta$         & $c_1$      & $c_3$ & $r_0/a$ \\  \hline
  &&&\\[-2ex]
  8.45            & -0.154846  & -0.0134070 & 5.29(7) \\ [0.7ex]
  8.0\phantom{0}  & -0.169805  & -0.0163414 & 3.69(4)
  \end{tabular}
\label{tabc}
\ee
In~(\ref{tabc}) we also quote the corresponding force parameters $r_0/a$,
as given in~\cite{Gattringer:2001jf}.
Assuming that $r_0=0.5\,\mbox{fm}$, they
translate into a lattice spacing of $a=0.095\,\mbox{fm}$ at $\beta=8.45$ and
$a=0.136\,\mbox{fm}$ at $\beta=8.0$. The mass parameter was chosen to be
$\rho=1.4$. This appeared to be a fair compromise between optimising the
condition number of $X^\dagger X$ as well as the locality properties of
$D_N$~\cite{Galletly:2005aa}.

The paper is organised as follows. In Section 2 we give a brief outline of our
calculations and present results for the renormalisation constants in one-loop
perturbation theory. In Section 3 we tadpole improve our results, and in
Section 4 we give our conclusions.

\section{Outline of the calculation and one-loop results}
\label{sec:2}

The Feynman rules specific for overlap fermions~\cite{ky,ikny} are collected
in~\cite{Horsley:2004mx}, while the gluon-operator and the gluon-gluon-operator
vertices (needed for the cockscomb and operator tadpole diagrams)
are independent of the fermion action and can be found in~\cite{Gockeler:1996hg}.
We consider general
covariant gauges, specified by the gauge parameter $\xi$. The Landau
gauge corresponds to $\xi=1$, while the Feynman gauge corresponds to
$\xi=0$.
In lattice momentum space the gluon propagator $D_{\mu\nu}(k)$ is given by the
set of linear equations
\be
\sum_\rho \left[G_{\mu\rho}(k) -
  \frac{\xi}{\xi-1}\hat{k}_\mu\hat{k}_\rho\right]  D_{\rho\nu}(k) =
        \delta_{\mu\nu} \, ,
\label{ortho}
\ee
where
\be
G_{\mu\nu}(k) = \hat{k}_\mu\hat{k}_\nu + \sum_\rho \left(
\hat{k}_\rho^2 \delta_{\mu\nu} - \hat{k}_\mu\hat{k}_\rho \delta_{\rho\nu}
\right)  \, d_{\mu\rho}
\ee
and
\be
d_{\mu\nu}= \left(1-\delta_{\mu\nu}\right)
\left[C_0 -
C_1 \, a^2 \hat{k}^2 -  C_2 \, a^2( \hat{k}_\mu^2 + \hat{k}_\nu^2)
\right]
\,, \quad
\hat{k}_\mu = \frac{2}{a}\sin\frac{ak_\mu}{2}\,, \quad
        \hat{k}^2 = \sum_\mu \hat{k}_\mu^2  \,.
\label{abbrev}
\ee
The coefficients $\{C_i\}$ are related to the coefficients $\{c_i\}$ of the
improved action by
\be
 C_0 = c_0 + 8 c_1 + 8 c_3 \,, \,\,\,
C_1 = c_3\,, \,\,\, C_2 = c_1 - c_3 \,.
\label{C1-C2}
\ee
The calculations are done analytically
as far as this is possible using {\it Mathematica}.
Part of the numerical results have been checked by an independent routine.

The bare lattice operators $\mathcal{O}(a)$ are, in general, divergent as $a
\rightarrow 0$. We define finite renormalised operators by
\be
\mathcal{O}^{\mathcal{S}}(\mu) = Z_{\mathcal{O}}^{\mathcal{S}}(a,\mu)\,
\mathcal{O}(a)\,,
\ee
where $\mathcal{S}$ denotes the renormalisation scheme. We have assumed that
the operators do not mix under renormalisation, which is the case for the
operators considered in this letter. The renormalisation constants
$Z_{\mathcal{O}}$ are often determined in the $MOM$ scheme first
from the gauge fixed quark propagator $S_N$ and the amputated
Green function
$\Lambda_{\mathcal{O}}$ of the operator $\mathcal{O}$:
\bea
Z_\psi^{MOM}(a,\mu)\; S_N\; \big|_{p^2=\mu^2}
&=&  S^{\rm tree} \,, \\[0.3em]
\frac{Z^{MOM}_{\cal O}(a,\mu) }{Z^{MOM}_\psi(a,\mu)}\;
\Lambda_{\cal O}\; \big|_{p^2=\mu^2}
&=& \Lambda_{\cal O}^{\rm tree}
+{\rm \ other\  Dirac\  structures}\,.
\label{ZOlat}
\eea
(Note that $Z_\psi = 1/Z_2$.) The renormalisation constants can be converted
to the $\MS$ scheme,
\be
\begin{split}
Z_{\psi}^{\MS}(a,\mu) &=  Z_{\psi}^{\MS,MOM}
Z_{\psi}^{MOM}(a,\mu)\,, \\
Z_{\mathcal{O}}^{\MS}(a,\mu) &=  Z_{\mathcal{O}}^{\MS,MOM}
Z_{\mathcal{O}}^{MOM}(a,\mu)\,,
\end{split}
\ee
where $Z_{\psi}^{\MS,MOM}$, $Z_{\mathcal{O}}^{\MS,MOM}$ are calculable in
continuum perturbation theory, and therefore
are independent of the particular choice of lattice
gauge and fermion actions.

In~\cite{Horsley:2004mx}
the wave function renormalisation constants where found to be
\be
 Z_\psi^{MOM}(a,\mu) 
 = 1 -\frac{g^2 C_F}{16\pi^2} \,
\left[ 2 (1-\xi) \log (a \mu)+ 4.79201 \, \xi + b_\Sigma \right]
\label{zpsimom}
\ee
in the $MOM$ scheme,
and
\be
 Z_\psi^{\MS}(a,\mu)  = 1 -\frac{g^2 C_F}{16\pi^2} \,
\left[ 2 (1-\xi) \log (a \mu)+ 3.79201 \, \xi + b_\Sigma +1\right]
\label{zpsi}
\ee
in the $\MS$ scheme, with $C_F = 4/3$ and
\be
\begin{tabular}{c|c}
          $\beta$          & $b_\Sigma$ \\ \hline &  \\[-2ex]
            8.45           & -17.429    \\[0.7ex]
            8.0\phantom{0} & -17.054
\end{tabular}
\ee

\begin{figure}[!htb]
\begin{center}
\begin{tabular}{cc}
\includegraphics[scale=0.60,clip=true]{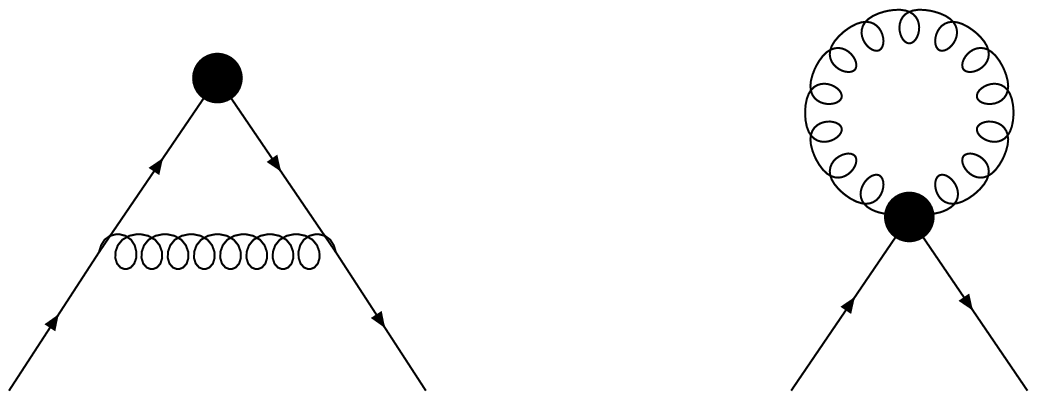} \hspace{1.5cm} &
\includegraphics[scale=0.60,clip=true]{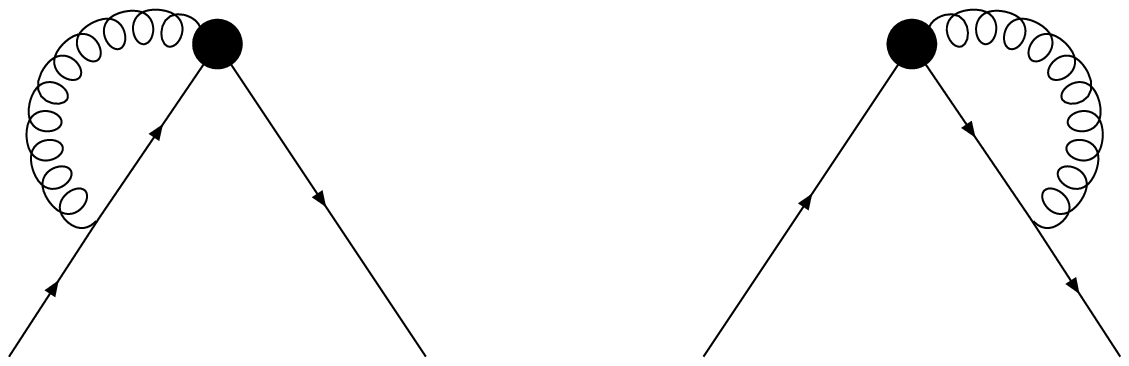}
\end{tabular}
\end{center}
\caption{The one-loop lattice Feynman diagrams contributing to
the amputated Green function. From left to right:  vertex, operator tadpole and
cockscomb diagrams.}
  \label{diags}
\end{figure}

We consider the following one-link operators
\bea
{\cal O}_{\mu\nu} &=& \frac{i}{2}\,\bar{\psi} (x) \gamma_\mu \Dlr_\nu
\psi(x) \,,
\label{onelinkOp}
\\
{\cal O}_{\mu\nu}^5 &=& \frac{i}{2}\,\bar{\psi} (x) \gamma_\mu\gamma_5 \Dlr_\nu
  \psi(x) \,,
  \label{onelinkOp5}
\eea
where $\Dlr_\nu=\Dr_\nu-\Dl_\nu$ is the (symmetric)  lattice covariant
derivative. While in our previous work~\cite{Horsley:2004mx},
which involved local bilinear quark operators, we only had to deal with the
vertex diagram shown on the left-hand side of Fig.~\ref{diags}, we now obtain contributions
from additional diagrams: the operator tadpole and the cockscomb diagrams shown on the
right-hand side of Fig.~\ref{diags}.

The amputated Green function of the operator ${\cal O}_{\mu\nu}$ [eq. (\ref{onelinkOp})]
turns out to be
\bea
\label{fullstructure}
\Lambda_{\mu\nu} (a,p) &=&
\gamma_\mu p_\nu +
\frac{g^2 C_F}{16\pi^2}\bigg\{
 \left[ \left(\frac{1}{3} +\xi  \right) \log (a^2 p^2)  - 4.29201 \, \xi +b_1 \right]
\gamma_\mu p_\nu
\nonumber
\\
&&\hspace{-6mm}
+\left[ \frac{4}{3}  \log (a^2 p^2) + \frac{1}{2} \,  \xi +b_2 \right]
\gamma_\nu p_\mu
+\left[- \frac{2}{3}  \log (a^2 p^2) - \frac{1}{2} \,  \xi +b_3 \right]
\delta_{\mu\nu} \pslash
\\
&&\hspace{-6mm}
+\,b_4 \, \delta_{\mu\nu} \gamma_\nu p_\nu  +\left( -\frac{4}{3}+\xi\right) \frac{p_\mu p_\nu}{p^2} \pslash
\bigg\}\,,
\nonumber
\eea
where $p$ is the external quark momentum,
and the coefficients $\{b_i\}$ are given in Table~\ref{bcoeff} for the
tadpole improved L\"uscher-Weisz action and, for comparison, for the
plaquette action (with $c_1=c_3=0$) as well.
\begin{table}[!t]
\begin{center}
\begin{tabular}{c|c|c|c|c}
Action  & $b_1$ & $b_2$& $b_3$& $b_4$ \\
  \hline
&&&&  \\[-2ex]
$\beta=8.45$ & -\phantom{0}5.6115  & -3.8336 & 2.7793 &  0.3446\\ [0.7ex]
$\beta=8.0\phantom{0}$ & -\phantom{0}5.2883 & -3.7636& 2.7310& 0.3331\\ [0.7ex]
Plaquette  & -10.6882   &  -4.7977  &  3.4612  &   0.5267
\end{tabular}
\end{center}
\caption{The coefficients $\{b_i\}$ for the tadpole improved L\"uscher-Weisz action at $\beta=8.45$ and
$8.0$, as well as for plaquette action.}
  \label{bcoeff}
\end{table}
The latter numbers are independent of $\beta$.
The  Green function $\Lambda_{\mu\nu}^5(a,p)$ of the operator ${\cal O}_{\mu\nu}^5$
[eq.~(\ref{onelinkOp5})] is obtained by multiplying
the right-hand side of (\ref{fullstructure}) by $\gamma_5$ from the right.
The coefficients $\{b_i^5\}$ turn out to be identical to $\{b_i\}$,
as is expected for overlap fermions. Thus, ${\cal O}_{\mu\nu}$ and ${\cal O}_{\mu\nu}^5$
have the same renormalisation constants. In the following we may therefore
restrict ourselves to the operator ${\cal O}_{\mu\nu}$.

It has been checked numerically
that the gauge dependent part  of (\ref{fullstructure}) is
universal ({\it i.e.} independent of the lattice gauge  and fermion action),
in accordance with the arguments presented in~\cite{Horsley:2004mx}.

Under the hypercubic group $H(4)$ the 16 operators of type (\ref{onelinkOp}) fall
into the following four irreducible representations~\cite{Gockeler:1996mu}:
\bea
\tau_3^{(6)}: \quad {\cal O}_{v_{2a}} & \equiv & \frac{1}{2}
\left({\cal O}_{14}+ {\cal O}_{41}  \right) \,,
\label{Ov2a} \\
\tau_1^{(3)}: \quad {\cal O}_{v_{2b}} & \equiv & {\cal O}_{44} -\frac{1}{3}
\left({\cal O}_{11}+ {\cal O}_{22}+{\cal O}_{33}  \right)\,,
\label{Ov2b}\\
\tau_1^{(1)}: \quad {\cal O}_{v_{2c}} & \equiv &
{\cal O}_{11}+ {\cal O}_{22}+{\cal O}_{33} +{\cal O}_{44}\,, \label{Ov2c} \\
\tau_1^{(6)}: \quad {\cal O}_{v_{2d}} & \equiv & {\cal O}_{14} - {\cal O}_{41}\,.
\label{Ov2d}
\eea
(We have given one example operator in each representation. A complete basis
for each representation can be found in~\cite{Gockeler:1996mu}.)
The operators (\ref{Ov2a}) and (\ref{Ov2b}) are widely used in
numerical simulations~\cite{Gockeler:1995wg, Gockeler:2004wp, Galletly:2003vf, Gurtler:2004ac}.
They correspond to the first moment of the parton distribution.
The operators (\ref{Ov2c}) and (\ref{Ov2d}) represent higher twist contributions
in the operator product expansion, and so are not used as much as operators
in the first two representations. For completeness we give results for all
four representations, so that the renormalisation factors for all operators
of the form (\ref{onelinkOp}) will be known.
We denote the corresponding amputated Green functions
by $\Lambda_{v_{2a}}$, $\Lambda_{v_{2b}}$, $\Lambda_{v_{2c}}$ and
$\Lambda_{v_{2d}}$. From (\ref{fullstructure}) we read off
\bea
\Lambda_{v_{2a}}\!\! &=&\! \frac{1}{2} \Big(\gamma_1 p_4 + \gamma_4 p_1\Big) \,\left\{1 +\frac{g^2
 C_F}{16\pi^2}
 \left[(\xi+\frac{5}{3}) \, \log(a^2 p^2) - 3.79201 \,\xi + b_{v_{2a}}
 \right] \right\}
\nonumber \\
&&\!  +\frac{g^2C_F}{16\pi^2} \left(-\frac{4}{3}+ \xi  \right)
\frac{p_1 p_4}{p^2} \pslash \,,
 \label{3PFa} \\
\Lambda_{v_{2b}}\!\! &=&\! \Big(\gamma_4 p_4 -\frac{1}{3}\sum_{i=1}^3
 \gamma_i p_i
 \Big) \,\left\{1 +\frac{g^2 C_F}{16\pi^2}
 \left[(\xi+\frac{5}{3}) \, \log(a^2 p^2) - 3.79201\, \xi + b_{v_{2b}}
 \right] \right\}
\nonumber \\
&&\!  +\frac{g^2C_F}{16\pi^2} \left(-\frac{4}{3}+ \xi  \right)
\left(p_4^2- \frac{1}{3}\sum_{i=1}^3 p_i^2\right)\frac{\pslash}{p^2}
 \,,
 \label{3PFb}\\
 \Lambda_{v_{2c}}\!\! &=&\! \pslash\,\left\{1 +\frac{g^2 C_F}{16\pi^2}
 \left[(\xi-1) \, \log(a^2 p^2) - 4.79201 \,\xi + b_{v_{2c}}
 \right] \right\}\,, \label{3PFc}\\
 \Lambda_{v_{2d}}\!\! &=&\! \Big(\gamma_1 p_4 - \gamma_4 p_1\Big) \,\left\{1 +\frac{g^2 C_F}{16\pi^2}
 \left[(\xi-1) \, \log(a^2 p^2) - 4.79201\, \xi + b_{v_{2d}}
 \right] \right\} \label{3PFd}
\eea
with
\be
b_{v_{2a}} = b_1+b_2\,,\quad b_{v_{2b}} = b_1+b_2+b_4\,,
\quad b_{v_{2c}} = b_1+b_2+4\,b_3 + b_4 -\frac{4}{3}\,,
\quad b_{v_{2d}} = b_1-b_2
\,.
\label{tabbv}
\ee
It is worth pointing out that with Wilson or clover fermions the Green functions
$\Lambda_{v_{2c}}$ and $\Lambda_{v_{2d}}$ both show perturbative mixing
of $O(g^2/a)$ with local operators. With overlap fermions these $O(1/a)$
terms are completely absent, showing once again that overlap fermions
behave much more like continuum fermions when mixing is a possibility.

Using (\ref{ZOlat}) and (\ref{zpsimom}), we obtain the renormalisation
constants in the $MOM$ scheme:
\bea
Z_{v_{2a},v_{2b}}^{MOM}(a,\mu)  &=&
  1 - \frac{g^2 C_F}{16 \pi^2} \left[\frac{16}{3} \log(a\mu) + \xi +
  b_{v_{2a},v_{2b}} + b_\Sigma
  \right]  \,, \\
Z_{v_{2c},v_{2d}}^{MOM}(a,\mu)  &=&
  1 - \frac{g^2 C_F}{16 \pi^2} \left[\,b_{v_{2c},v_{2d}} + b_\Sigma\,\right]  \,.
\eea
As already mentioned, the conversion factors $Z_{v_{2a},v_{2b},v_{2c},v_{2d}}^{\MS,MOM}$ are
universal~\cite{Gockeler:1996hg}. They are given by
\bea
Z_{v_{2a},v_{2b}}^{\MS,MOM}  &=&  1 - \frac{g^2 C_F}{16 \pi^2} \, \left(\frac{40}{9}-\xi\right)\,,
\\
Z_{v_{2c}}^{\MS,MOM}  &=&  1 - \frac{g^2 C_F}{16 \pi^2} \, \left(-\frac{4}{3}\right)\,,
\\
Z_{v_{2d}}^{\MS,MOM}  &=&  1 \,.
\eea
In the $\MS$ scheme we then find
\bea
Z_{v_{2a},v_{2b}}^{\MS}(a,\mu) & =&   1 - \frac{g^2 C_F}{16 \pi^2}
  \left[\frac{16}{3}
  \log(a\mu)+ \frac{40}{9} + b_{v_{2a},v_{2b}} + b_\Sigma  \right]  \, ,\\
Z_{v_{2c}}^{\MS}(a,\mu) & = &  1 - \frac{g^2 C_F}{16 \pi^2}
  \left[ -\frac{4}{3} + b_{v_{2c}} + b_\Sigma  \right]  \, ,\\
Z_{v_{2d}}^{\MS}(a,\mu)  &=&   1 - \frac{g^2 C_F}{16 \pi^2}\left[ b_{v_{2d}} + b_\Sigma  \right]  \,.
\eea

\section{Tadpole improved results}

A detailed discussion of mean field -- or tadpole -- improvement for overlap fermions
and extended gauge actions has been given in~\cite{Horsley:2004mx}. Here we
will briefly recall the basic idea, before presenting our results.

Tadpole improved renormalisation constants are defined by
\be
Z_\mathcal{O}^{TI} = Z_\mathcal{O}^{MF}
\left(\frac{Z_\mathcal{O}}{Z_\mathcal{O}^{MF}}\right)_{\rm pert}\,,
\label{zti}
\ee
where $Z_\mathcal{O}^{MF}$ is the mean field approximation of
$Z_\mathcal{O}$, while the right-hand factor is computed in perturbation
theory.
For overlap fermions (with $r=1$), and operators with $n_D$ covariant derivatives,
we have
\be
Z_{\cal O}^{MF} =  \frac{\rho \, u_0^{1-n_D} }{\rho -4 (1- u_0)} \,.
\label{ZOMF}
\ee
In our case
$n_D=1$. It is required that $\rho > 4(1-u_0)$, which is fulfilled here
(see Table~\ref{tabku}).
\begin{table}[!htb]
\begin{center}
  \begin{tabular}{c|c|c}
 $\beta$ &  $k_u^{TI}$ & $u_0^4$ \\ \hline
 &&\\[-2ex]
 8.45 &  0.543338\,$\pi^2$ &  0.65176 \\[0.7ex]
 8.0\phantom{0} & 0.515069\,$\pi^2$ &  0.62107
  \end{tabular}
  \end{center}
  \caption{The coefficient $k_u^{TI}$ and the average plaquette
 $u_0^4$ at $\beta=8.45$ and $8.0$.}
  \label{tabku}
\end{table}

To compute the right-hand factor in (\ref{zti}), we have to remove the
tadpole contributions from the perturbative expressions of $Z_{\cal O}$ first.
This is achieved if we
re-express the perturbative series in terms of tadpole improved
coefficients:
\bea
\frac{ c_0^{TI} }{ g_{TI}^2 } = u_0^4 \; \frac{c_0 }{ g^2 }\,, \quad \quad
\frac{ c_i^{TI} }{ g_{TI}^2 } = u_0^6 \; \frac{c_i }{ g^2 }
\,, \quad i=1, 3\,.
\eea
This does not fix all parameters, but leaves us with some freedom of
choice. The simplest choice is to define
\bea
g_{TI}^2 =\frac{g^2}{u_0^4} \,,\quad  \quad
c_0^{TI} = c_0 \,, \quad  \quad
c_i^{TI} = u_0^2 \,c_i \,, \quad i=1, 3\,.
\label{cti}
\eea
With this choice
 \be
 C_0^{TI} = c_0 + 8  c_1^{TI} + 8  c_3^{TI} \,,
 \quad
 C_1^{TI} = u_0^2\, C_1\,, \quad C_2^{TI} = u_0^2\, C_2\,.
 \ee
(Note that $C_0^{TI} \neq 1$. However, $C_0^{TI}
\rightarrow 1$ in the continuum limit.) This means that we have to replace every
$g^2$ by $g_{TI}^2$ and every $c_1$ and $c_3$ by $c_1^{TI}$ and $c_3^{TI}$,
respectively, while keeping $c_0$ unchanged. The effect
of introducing tadpole improved coefficients (\ref{cti}) is that the rescaled
gluon propagator remains of the same form as we change $u_0$, thus ensuring
fast convergence.

To compute $Z_{\cal O}^{MF}$ perturbatively, we need to know the perturbative
expansion of $u_0$ to one-loop order~\cite{Lepage:1992xa,Alford:1995hw}.
We write
\be
u_0 = 1 - \frac{g_{TI}^2 C_F}{16 \pi^2}\, k_u^{TI}  \,.
\label{exu}
\ee
In \cite{Horsley:2004mx} we have computed $k_u^{TI}$ for the L\"uscher-Weisz
action with coefficients $C_0^{TI}$, $C_1^{TI}$ and $C_2^{TI}$. The numbers
are given in Table \ref{tabku} for our two values of $\beta$, 
together with the `measured' values of $u_0^4$. Expanding (\ref{ZOMF}) then gives
\be
 Z_{\cal O\,{\rm pert}}^{MF} =  1 + \frac{g_{TI}^2 C_F}{16 \pi^2} \,
 \frac{4}{\rho} k_u^{TI}\,.
 \label{ZOmfpert}
\ee

Let us now rewrite the one-loop renormalisation constants of Section~\ref{sec:2}
as
\bea
Z_{v_{2a},v_{2b}} &=&  1 - \frac{C_F\, g^2}{16 \pi^2}
 \left[ \frac{16}{3\,C_0} \log(a\mu) + B_{v_{2a},v_{2b}}(\rho,C) \right] \, ,
\label{BOdefab}
\\
Z_{v_{2c},v_{2d}} &=&  1 - \frac{C_F\, g^2}{16 \pi^2}
B_{v_{2c},v_{2d}}(\rho,C)  \, .
\label{BOdefcd}
\eea
Dividing (\ref{BOdefab}) and (\ref{BOdefcd}) by (\ref{ZOmfpert}) and inserting
(\ref{ZOMF}), we obtain mean field/tadpole improved
renormalisation constants:
\bea
Z_{v_{2a},v_{2b}}^{TI} &=& \frac{\rho}{\rho-4(1-u_0)}
 \left\{1 - \frac{g_{TI}^2 C_F}{16 \pi^2} \left[
  \frac{16}{3\,C^{TI}_0} \log(a\mu)
  + B_{v_{2a},v_{2b}}^{TI} \right]\right\} \,,
\label{ZTIab}
\\
Z_{v_{2c},v_{2d}}^{TI} &=& \frac{\rho}{\rho-4(1-u_0)}
 \left\{1 - \frac{g_{TI}^2 C_F}{16 \pi^2}  B_{v_{2c},v_{2d}}^{TI}\right\}\,,
\label{ZTIcd}
\eea
where we have introduced the abbreviated notation
\be
B^{TI} = B(\rho,C^{TI})
 + \frac{4}{\rho}\,k_u^{TI}\,.
\ee
The coefficients $B(\rho,C^{TI})$ are the analogue of
$B(\rho,C)$, with $C_0$, $C_1$ and $C_2$ being
 replaced by $C_0^{TI}$, $C_1^{TI}$
and $C_2^{TI}$, respectively.
In (\ref{ZTIab}) and (\ref{ZTIcd}) only the gluon propagator
has been tadpole improved.

To tadpole improved the fermion propagator as well, we must replace $\rho$
by~\cite{Horsley:2004mx}
\be
\rho^{TI} = \frac{\rho-4(1-u_0)}{u_0}
\label{rhoti}
\ee
in the right-hand perturbative factor of (\ref{zti}).
This defines `fully tadpole improved' renormalisation constants
\bea
Z_{v_{2a},v_{2b}}^{FTI} &=& \frac{\rho}{\rho-4(1-u_0)}
 \left\{1 - \frac{g_{TI}^2 C_F}{16 \pi^2} \left[
  \frac{16}{3\,C^{TI}_0} \log(a\mu)
  + B_{v_{2a},v_{2b}}^{FTI} \right]\right\} \,,
\\
Z_{v_{2c},v_{2d}}^{FTI} &=& \frac{\rho}{\rho-4(1-u_0)}
 \left\{1 - \frac{g_{TI}^2 C_F}{16 \pi^2}  B_{v_{2c},v_{2d}}^{FTI}\right\}
\eea
with
\be
B^{FTI} = B(\rho^{TI},C^{TI})+ \frac{4}{\rho^{TI}}\,k_u^{TI}\,.
\ee

In Table~\ref{ZSTab} we present our final results and compare tadpole improved
and unimproved renormalisation constants.
\begin{table}[!tb]
\begin{center}
\begin{tabular}{c|c|c|c|c|c|r|c}
Operator  & $\beta$ & $B$ & {$Z^\MS$}&
{$B^{TI}$}& {$Z^{TI,\MS}$} &
{$B^{FTI}$}& {$Z^{FTI,\MS}$}\\
  \hline
  &&&&&&&\\[-2ex]
  $v_{2a}$    &  $8.45$         & $-22.430$  &$1.315$ &
                $0.502$
                &$1.393$        &$-0.077$ & $1.411$\\[0.7ex]
  $v_{2b}$    &  $8.45$         & $-22.085$  &$1.311$ &
                $0.793$
                &$1.384$        &$0.230$ & $1.401$\\[0.7ex]
  $v_{2c}$    &  $8.45$         & $-18.079$  &$1.254$ &
                $2.985$
                &$1.318$        &$1.829$ & $1.353$\\[0.7ex]
  $v_{2d}$    &  $8.45$         & $-19.207$  &$1.270$ &
                $2.303$
                &$1.338$        &$1.369$ & $1.367$\\[0.7ex]
                \hline
&&&&&&&\\[-2ex]
  $v_{2a}$    &  8.0\phantom{0} & $-22.036$
                &$1.310$        & $0.603$
                &$1.390$        &$-0.108$ & $1.412$ \\[0.7ex]
  $v_{2b}$    &  8.0\phantom{0} & $-21.703$
                &$1.305$        & $0.892$
                &$1.381$        &$0.199$ & $1.402$  \\[0.7ex]
  $v_{2c}$    &  8.0\phantom{0} & $-17.890$
                &$1.252$        & $3.038$
                &$1.316$        &$1.643$ & $1.358$ \\[0.7ex]
  $v_{2d}$    &  8.0\phantom{0} & $-18.954$
                &$1.266$        & $2.371$
                &$1.336$        &$1.239$ & $1.371$
\end{tabular}
\end{center}
\caption{The constants $B$ and $Z^\MS$ at $a=1/\mu$ for various levels of improvement.}
  \label{ZSTab}
\end{table}
We see that the improved coefficients $B$ are rather small in the case
of the operators $v_{2a}$ and $v_{2b}$, much smaller than for Wilson
and clover fermions~\cite{Capitani:2001xi}, which raises hope that
the perturbative series converges rapidly. This furthermore means
that the dominant contribution to the renormalisation constants is
given by the mean field factor (\ref{ZOMF}).

\section{Summary}

We have computed the renormalisation constants of one-link quark operators for
overlap fermions and tadpole improved L\"uscher-Weisz action for two values of
the coupling, $\beta=8.45$ and $8.0$, being used in current simulations. The
calculations have been performed in general covariant gauge, using
the symbolic language {\it Mathematica}. This gave us complete control over
the Lorentz and spin structure, the cancellation of infrared divergences, as
well as the cancellation of $1/a$ singularities. However, the price is
high. In intermediate steps we had to deal with $O(10^5)$ terms due to the
complexity of the gauge field action.

To improve the convergence of the perturbative series and to get rid of lattice
artefacts, we have applied tadpole improvement to our results. This was done
in two stages. In the first stage we improved the gluon propagator, while in
the second stage we improved both gluon and quark propagators.

Results at other $\beta$ values, $\rho$ parameters
(also including other gauge field actions with up to six links)
can be provided on request.

\section*{Acknowledgements}

This work is supported by DFG under contract FOR 465 (Forschergruppe
Gitter-Hadronen-Ph\"{a}nomenologie) and by the EU Integrated Infrastructure
Initiative Hadron Physics (I3HP) under contract RII3-CT-2004-506078.

\end{document}